\documentclass[journal=nalefd,manuscript=letter,layout=twocolumn]{achemso}
\setkeys{acs}{articletitle = true} 

\usepackage[T1]{fontenc} 
\usepackage{graphicx}
\usepackage[absolute]{textpos}

\newcommand*{\VB}{V\textsubscript{B}\textsuperscript{-}}
\newcommand*{\VN}{V\textsubscript{N}}
\newcommand*{\VBz}{V\textsubscript{B}\textsuperscript{0}}
\newcommand*{\VBn}{V\textsubscript{B}}
\DeclareUnicodeCharacter{2212}{\textendash}
\author{David Curie}
\affiliation[Vanderbilt University]
{Department of Physics and Astronomy, Vanderbilt University, Nashville, TN, USA}
\author{Jaron T. Krogel}
\affiliation[ORNL]
{Materials Science and Technology Division, Oak Ridge National Laboratory, Oak Ridge, TN, 37831, USA}
\author{Lukas Cavar}
\affiliation[Indiana University]
{Indiana University, Bloomington, IN, USA}
\author{Abhishek Solanki}
\affiliation[Purdue University]
{Elmore Family School of Electrical and Computer Engineering, Purdue University, West Lafayette, IN, 47907 USA}
\author{Pramey Upadhyaya}
\affiliation[Purdue University]
{Elmore Family School of Electrical and Computer Engineering, Purdue University, West Lafayette, IN, 47907 USA}
\author{Tongcang Li}
\affiliation[Purdue University]
{Elmore Family School of Electrical and Computer Engineering, Purdue University, West Lafayette, IN, 47907 USA}
\author{Yun-Yi Pai}
\affiliation[ORNL]
{Materials Science and Technology Division, Oak Ridge National Laboratory, Oak Ridge, TN, 37831, USA}
\alsoaffiliation{Quantum Science Center, Oak Ridge, Tennessee 37831, USA}
\author{Michael Chilcote}
\affiliation[ORNL]
{Materials Science and Technology Division, Oak Ridge National Laboratory, Oak Ridge, TN, 37831, USA}
\alsoaffiliation{Quantum Science Center, Oak Ridge, Tennessee 37831, USA}
\author{Vasudevan Iyer}
\affiliation[CNMS]
{Center for Nanophase Materials Sciences, Oak Ridge National Laboratory, Oak Ridge, TN, 37831, USA}
\author{Alex Puretzky}
\affiliation[CNMS]
{Center for Nanophase Materials Sciences, Oak Ridge National Laboratory, Oak Ridge, TN, 37831, USA}
\author{Ilia Ivanov}
\affiliation[CNMS]
{Center for Nanophase Materials Sciences, Oak Ridge National Laboratory, Oak Ridge, TN, 37831, USA}

\author{Mao-Hua Du}
\affiliation[ORNL]
{Materials Science and Technology Division, Oak Ridge National Laboratory, Oak Ridge, TN, 37831, USA}
\author{Fernando Reboredo}
\affiliation[ORNL]
{Materials Science and Technology Division, Oak Ridge National Laboratory, Oak Ridge, TN, 37831, USA}
\author{Benjamin Lawrie}
\affiliation[ORNL]
{Materials Science and Technology Division, Oak Ridge National Laboratory, Oak Ridge, TN, 37831, USA}
\alsoaffiliation{Quantum Science Center, Oak Ridge, Tennessee 37831, USA}
\email{lawriebj@ornl.gov}

\title{Correlative nanoscale imaging of strained hBN spin defects}

\begin{document}

\footnotetext{This manuscript has been authored by UT-Battelle, LLC, under contract DE-AC05-00OR22725 with the US Department of Energy (DOE). The US government retains and the publisher, by accepting the article for publication, acknowledges that the US government retains a nonexclusive, paid-up, irrevocable, worldwide license to publish or reproduce the published form of this manuscript, or allow others to do so, for US government purposes. DOE will provide public access to these results of federally sponsored research in accordance with the DOE Public Access Plan (http://energy.gov/downloads/doe-public-access-plan).}

\begin{abstract}
Spin defects like the negatively charged boron vacancy color center (\VB{}) in hexagonal boron nitride (hBN) may enable new forms of quantum sensing with near-surface defects in layered van der Waals heterostructures. Here, we reveal the effect of strain associated with creases in hBN flakes on \VB{} and \VBz{} color centers in hBN with correlative cathodoluminescence and photoluminescence microscopies. We observe strong localized enhancement and redshifting of the \VB{} luminescence at creases, consistent with density functional theory calculations showing \VB{} migration toward regions with moderate uniaxial compressive strain. The ability to manipulate these spin defects with highly localized strain offers intriguing possibilities for future 2D quantum sensors.
\end{abstract}


Color centers in hBN have drawn increasing interest since single-photon emission from hBN color centers was reported in 2016\cite{tran_robust_2016,aharonovich_solid-state_2016,tran2016quantum,PhysRevB.94.121405}. More recently, spin defects in hBN have drawn interest for applications in quantum sensing, quantum networking, and quantum computing\cite{stern2021room,PhysRevB.104.075410,gottscholl_room_2021,baber2021excited,gottscholl_initialization_2020,kianinia_generation_2020,mathur2021excited,murzakhanov2021electron,mu2021excited,tetienne2021quantum}. The development of new quantum technologies based on \VB{} spin defects in hBN depends on improved understanding and control of the effect of the hBN environment on the \VB{} luminescence.  Unfortunately, establishing links between atomic structure, nanoscale morphology, and mesoscale optical properties remains challenging. Transmission electron microscopes \cite{jin2009fabrication} and scanning tunneling microscopes \cite{qiu2021atomic,wong2015characterization} have been used to image and manipulate defects in hBN, but because (1) optically active defects are generally sparsely distributed, (2) high energy electron beams can modify atomic defects, and (3) optical access in these microscopes is generally limited, correlating atomic and nanoscale structure with optical behaviour remains a challenge. We have recently shown that multicolor photon correlation functions can be critical to understanding the correlations between different hBN PL bands\cite{feldman2021evidence,feldman2019phonon}, but additional research is needed to help accelerate basic research in quantum nanophotonics toward practical 2D quantum technologies.

\begin{figure*}[hbt!]
    \centering
    \includegraphics[width=1.9\columnwidth]{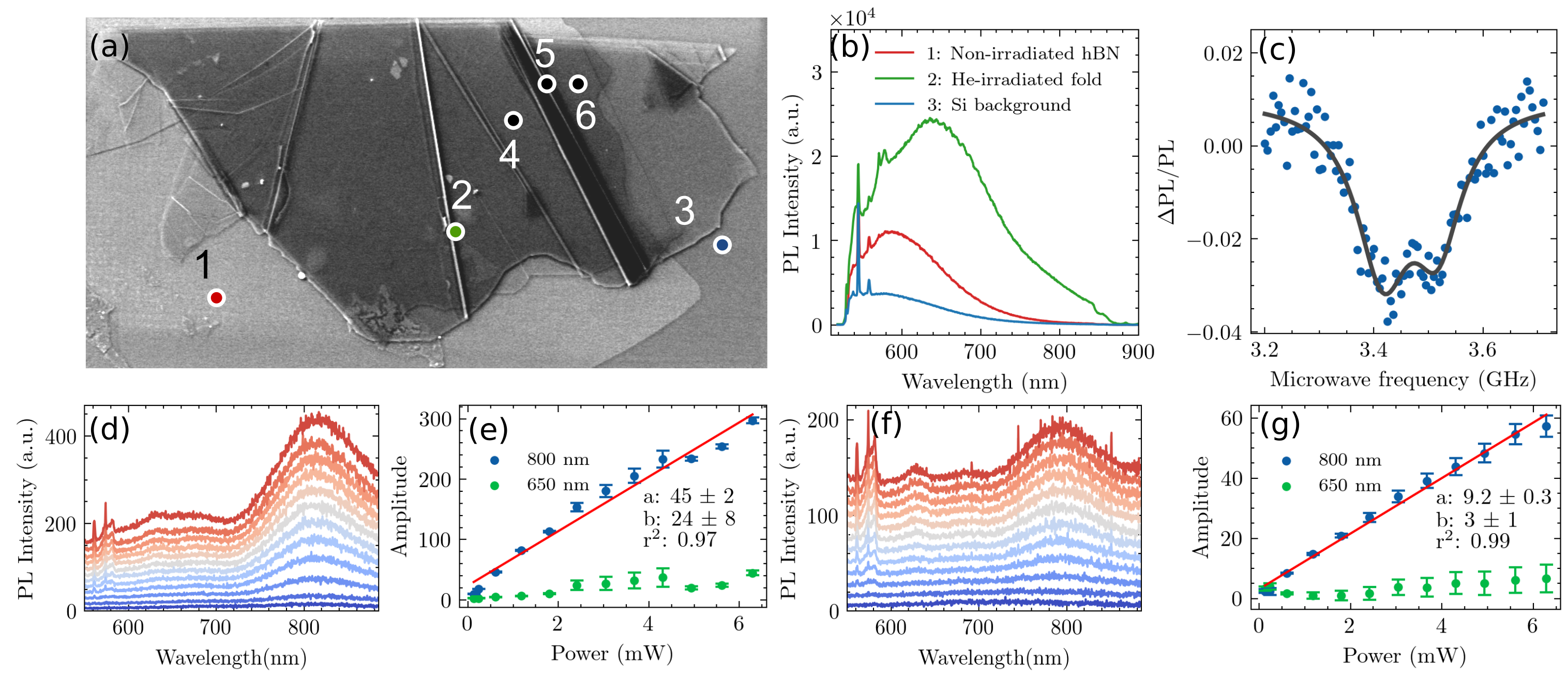}
    \caption{(a) 46.0~$\mu$m~x~23.7~$\mu$m SEM image of a He-irradiated hBN flake. (b) PL spectra at positions 1, 2, and 3. (c) continuous wave optically detected magnetic resonance (ODMR) spectrum acquired with a 700 nm long pass filter at position 4. (d,e) Power-dependent PL on a crease at position 5 with associated Lorentz fit parameters. (f,g) Power-dependent PL at position 6 with associated Lorentz fit parameters. In (d) and (f), the power increases monotonically from blue to red from 0.13 mW to 6.3 mW. In (e) and (g), the fitted amplitude of the \VB{} luminescence is shown in blue, the fitted amplitude of the visible luminescence is shown in green, and a linear fit to the \VB{} PL power dependence is included with fit parameters in the inset.  A constant 10-unit offset is used between spectra in the waterfall plots in (d) and (f).}
    \label{fig:summary_overview}
\end{figure*}




\VB{} color centers have been patterned with ion beams\cite{kianinia_generation_2020,gao_highcontrast_2021}, lasers\cite{gao2021femtosecond}, neutrons\cite{gottscholl_room_2021}, and electron beams \cite{murzakhanov2021creation}, and their photophysical properties are generally consistent with theory\cite{abdi_color_2018,libbi2021phonon,chen2021photophysical,li2020giant}, though control over the distribution of \VB{}, \VBz{}, and nitrogen vacancy (\VN{}) defects is still not well understood. While strain has proven to be a flexible tool for controlling single photon emitters in hBN\cite{hayee_revealing_2020,li2020giant,deng_strain_2018,mendelson_straininduced_2020,proscia2018near,yim_polarization_2020}, many questions remain about the effect of strain on \VB{} color centers.
Despite substantial recent progress, \VB{} PL exhibits poor quantum efficiency compared with more established optically accessible spin defects like the nitrogen vacancy center in diamond; there has still been no report of single \VB{} defect emission.  

Cathodoluminescence (CL) microscopies have found increasing relevance in the characterization of color centers in hBN with spatial resolution beyond the diffraction limit\cite{hayee_revealing_2020,bourrellier2016bright,meuret2015photon}. The electron beam in the scanning electron microscope (SEM) used here has a spot size of 0.7 nm in high vacuum that is increased to 2-3 nm at 5 keV in low vacuum. The CL spatial resolution is further constrained by free-carrier migration after electron-beam excitation, but CL still offers 1-2 orders of magnitude better spatial resolution than PL microscopy. CL also offer additional opportunities to probe excitation and relaxation pathways that are not present in PL spectrum images and to correlate nanoscale morphology with device-scale optical properties.
Unfortunately, electron-beam interactions with matter are typically weak in the thin-flake limit. Increased electron-beam doses can result in beam-induced damage, while reduced doses yield small signals that are challenging to interpret. Moreover, defects in oxide substrates often exhibit bright CL that can swamp the CL signal from 2D flakes. Recent work has shown that this background CL from the substrate can be used to quantify the transmission of the substrate CL through a flake\cite{negri_quantitative_2020}, though understanding the convolution of 2D material CL, substrate CL, and the affect of an exfoliated flake on the substrate CL can be challenging. Here, we reveal the effect of strain associated with creases in hBN flakes on \VB{}, \VBz{}, and \VN{} color centers in hBN with non-negative matrix factorization (NMF)\cite{NMF:review} of CL and PL spectrum images.

Ensemble \VB{}, \VBz{}, and \VN{} defects were introduced into a hBN flake ($\sim$ 10-20 nm thick) by helium ion irradiation\cite{kianinia_generation_2020}, and that flake was transferred onto an un-irradiated hBN flake (that serves as a reference system with no extrinsic defects) on a 285 nm thick SiO\textsubscript{2} layer on a Si substrate as shown in a SEM image in Fig. \ref{fig:summary_overview}(a). Ambient PL microscopies were performed using a 532 nm continuous wave laser excitation and a 100x objective.  CL microscopy was performed in a FEI Quattro environmental scanning electron microscope with a Delmic Sparc CL collection module.  The electron beam in the SEM excites the sample through a pinhole in a parabolic mirror.  The parabolic mirror collimates the resulting CL for characterization with an Andor Kymera spectrometer, and the secondary electron (SE) signal is concurrently measured on an Everhart-Thornley detector.  All CL spectrum images reported here were acquired at 5~keV in a 0.3 mbar water vapor background. The electron beam was rastered within each pixel of the CL image during its measurement dwell time to generate a high-pixel-density concurrent SEM image.

Several spectral features are immediately visible in hyperspectral PL maps of the flake and illustrated in point spectra shown in  Figs. \ref{fig:summary_overview}(b), (d), and (f). Figure \ref{fig:summary_overview}(b) illustrates typical PL point spectra from the bare substrate, from un-irradiated hBN, and from one position on helium-irradiated hBN. Silicon and hBN Raman lines and broad defect luminescence bands are visible in each. The \VB{} emission appears as a shoulder at wavelengths near 800 nm in the PL spectrum acquired at position 2, but it is not present as a distinct well-resolved peak at all positions on the flake. 

Increased \VB{} PL is seen at $\sim$ 750-900 nm at positions 5 and 6 in Fig. \ref{fig:summary_overview}(a). The power-dependent PL measured at those sites is plotted in Figs. \ref{fig:summary_overview}(d) and \ref{fig:summary_overview}(f) respectively. An ODMR spectrum acquired near position 4 and shown in Fig.~\ref{fig:summary_overview}(c) exhibits moderate contrast of almost 4\%, confirming the presence of ensemble \VB{} defects. The wavelength and amplitude of the PL spectra in Figs.~\ref{fig:summary_overview}(d) and ~\ref{fig:summary_overview}(f) were fitted to two Lorentzians; the fitted amplitude parameters are plotted as a function of excitation laser power in Figs. \ref{fig:summary_overview}(e) and (g) respectively with linear fits to the \VB{} power dependence included in each graph. 

\begin{figure*}[hbt!]
    \centering
    \includegraphics[width=\textwidth]{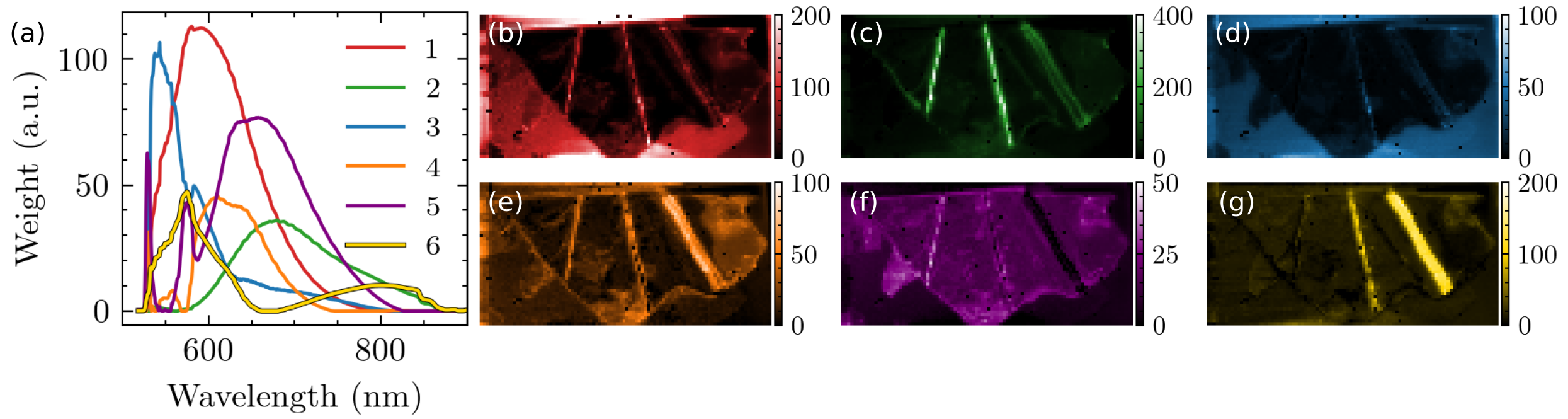}
    \caption{NMF decomposition of hyperspectral photoluminescence map of the same He-irradiated hBN flake shown in Figure 1. The spectral components illustrated in (a) are color coded to the six associated NMF maps in (b-g).}
    \label{fig:nmf_pl}
\end{figure*}

Notably, the relative intensity of the \VB{} PL is highly heterogeneous and the \VB{} energy redshifts by $\sim$ 30~nm in the presence of strain at the crease.  In addition, the \VB{} PL intensity grows much more quickly with increasing laser power than the visible PL, and the \VB{} PL intensity grows much more quickly on the crease than off the crease. Linear least-squares fits to the power-dependent PL amplitude plotted in Figs. \ref{fig:summary_overview}(e) and (g) show a slope that is $\sim 5\times$ larger for \VB{} PL on the crease than off the crease. The relatively weak signal at visible wavelengths at positions 5 and 6 limits the quality of the fit to the visible luminescence, but the difference in scaling of the visible and \VB{} PL on and off the crease is clear. The linewidths and central wavelengths of the fits showed no statistically significant dependence on excitation power.

The heterogeneity of the \VB{} PL is illustrated in Fig.~\ref{fig:nmf_pl}, which shows the NMF decomposition of a PL spectrum image of the same flake acquired with a 2.2 mW laser excitation. NMF is a popular, computationally inexpensive tool for extracting sparse, physically relevant data from hyperspectral datasets that assumes that a spectrum image results from the linear combination of non-negative constituent spectra\cite{iyer2021near,liu2019spatiotemporal}. PL component 1 in Figs.~\ref{fig:nmf_pl}(a) and ~\ref{fig:nmf_pl}(b) has a peak at 600~nm that is consistent with the un-irradiated background hBN PL spectrum in Fig.~\ref{fig:summary_overview}(b), and it exhibits a strong spatial correlation with the un-irradiated hBN. Likewise, PL component 3 in Figs.~\ref{fig:nmf_pl}(a) and ~\ref{fig:nmf_pl}(d) is consistent with the measured substrate luminescence.

\begin{figure*}[hbt!]
    \centering
    \includegraphics[width=\textwidth]{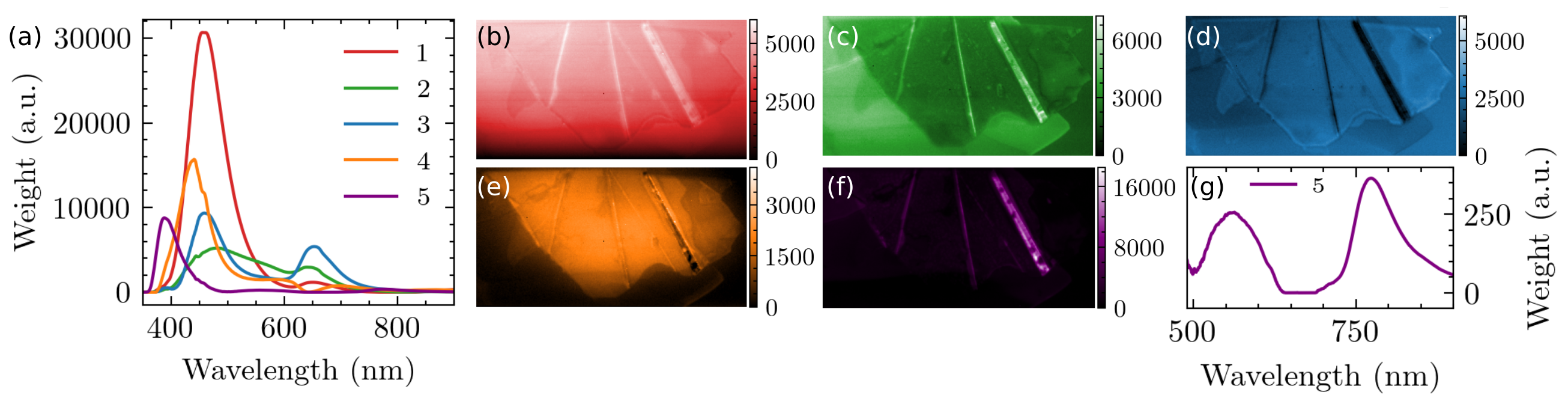}
    \caption{NMF decomposition of a hyperspectral cathodoluminescence map of the same He-irradiated hBN flake shown in Figure 1. (a) NMF spectral components corresponding to their color-coded spatial maps in (b--f). (g) NMF spectral component 5 from 500-900 nm (highlighting the weak emission at 560 nm and 790 nm).}
    \label{fig:nmf_cl}
\end{figure*}

There is a clear positive correlation between the creases and the amplitude of PL components 2, 4, and 6. PL component 6 comprises the \VB{} PL at 800 nm along with correlated visible emission near 560 nm, and it is strongly enhanced by two of the creases in the flake, as seen in Fig.~\ref{fig:nmf_pl}(g).  PL components 2 and 4 comprise visible emission bands that are also enhanced at creases in the flake as seen in Figs.~\ref{fig:nmf_pl}(c) and (e) respectively.  PL component 2 may include \VB{} PL in the shoulder of a higher energy peak, though it is not as well resolved as the \VB{} band in component 6. PL component 5 is strongly spatially correlated with the He-irradiated hBN flake, but it is attenuated at the creases, and it shows no emission at the \VB{} PL band near 800 nm. The origins of the spatial heterogeneity of the hBN PL is discussed below.

A rastered CL spectrum image of the same flake was acquired using a beam current of 1800~pA, a step size  of 400 nm, and a dwell time of 600 ms per pixel. The NMF decomposition of that CL spectrum image is illustrated in Fig. \ref{fig:nmf_cl}. CL component 1 and 3 in Figs.~\ref{fig:nmf_cl}(a), ~\ref{fig:nmf_cl}(b), and ~\ref{fig:nmf_cl}(d) are consistent with the substrate luminescence\cite{negri_quantitative_2020}, and CL component 2 in Figs.~\ref{fig:nmf_cl}(a) and ~\ref{fig:nmf_cl}(c) is spatially correlated with the un-irradiated flake, and can likely be attributed to intrinsic defect CL from that flake. Like PL component 5, CL component 4 is present across the He-irradiated hBN flake, but attenuated at the creases, and it exhibits negligible intensity near 800 nm. CL component~5 exhibits weak emission around 790~nm with additional emission at 560~nm and 400~nm, as seen in the magnified low energy portion of CL component~5 in Fig.~\ref{fig:nmf_cl}(g). It is strongly enhanced by one of the largest creases on the flake with weaker enhancement at the other creases. The correlation between emission at 790~nm and 560~nm  is consistent with the crease-enhanced luminescence in PL component 6.  The bright emission at 400 nm that is spatially correlated with the defect bands at 790~nm and 560~nm in CL component 5 has not been previously reported in the literature.

\begin{figure*}[hbt!]
    \centering
    \includegraphics[width=0.8\textwidth]{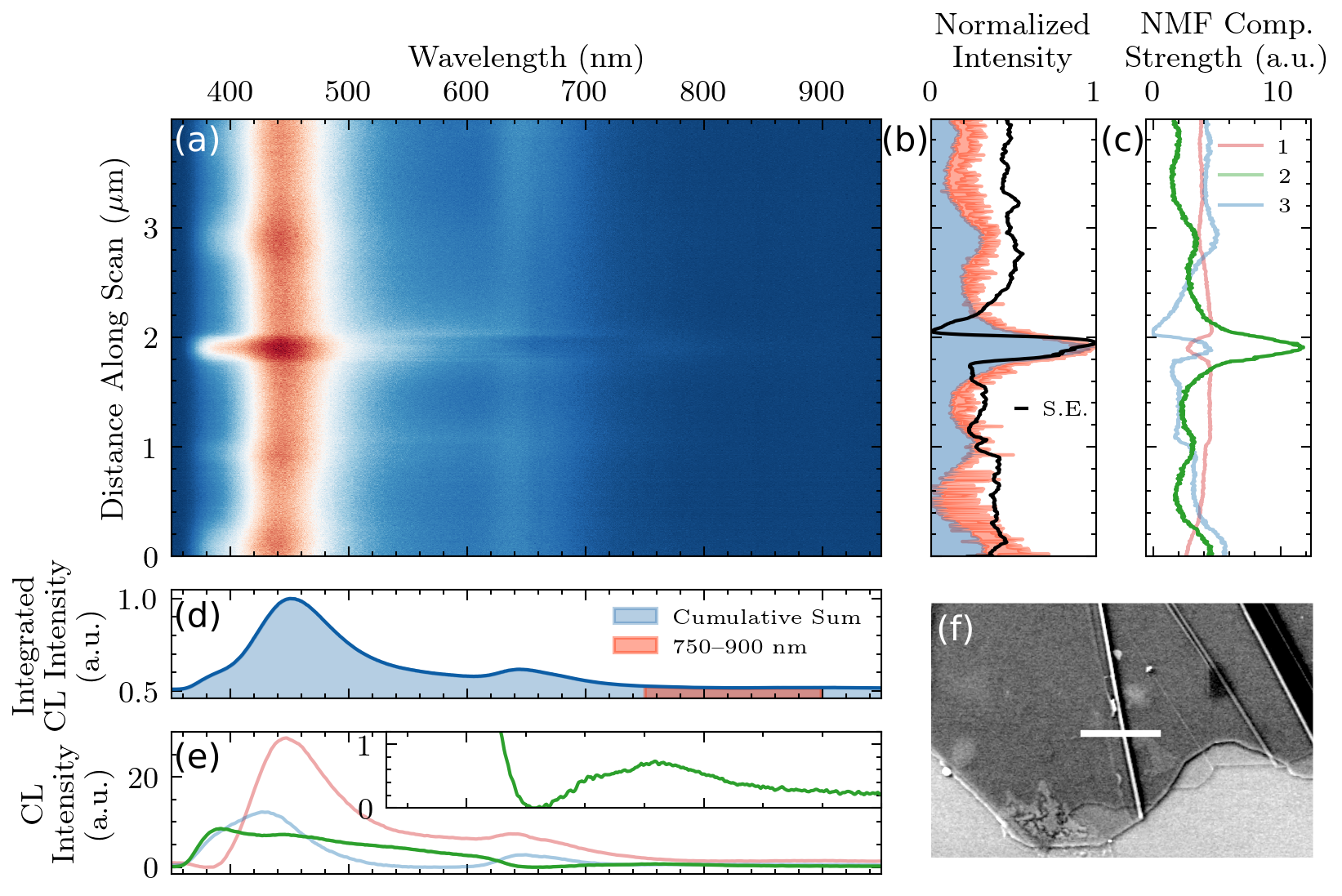}
    \caption{A 4 $\mu m$ long CL line scan (a) was acquired across the white line in the SEM image (f). (b) normalized panchromatic CL intensity as a function of linescan position (blue) with normalized SE counts (black) and the normalized counts across the \VB{} band (red). (d) integrated spectra across all linescan positions (blue) with the \VB{} band highlighted (red). (c,e) NMF decomposition of linescan with magnified low energy values for component 2 shown in the inset of (e).  Wavelength axis is shared in (a,d,e) and linescan distance axis is shared in (a,b,c).}
    \label{fig:linescan}
\end{figure*}

Because the total measurement time and electron-beam dose scale as the inverse square of the step size, spectrum images utilizing step sizes much smaller than 400 nm are problematic for flakes this large, but line scans that focus on individual features with higher spatial resolution can still be performed quickly. A CL line scan was acquired using a beam current of 1~nA, a 100~ms dwell time, and a 5~nm step size, as illustrated in Fig.~\ref{fig:linescan}. The spectrum image is dominated by a defect band near 450~nm, just as the spectrum image in Fig.~\ref{fig:nmf_cl} is. The crease is identified in the CL line-scan by the sudden enhancement in intensity in Figs.~\ref{fig:linescan}(a) and (b) and the increased SE signal in Fig.~\ref{fig:linescan}(b). However, it is not immediately clear from Fig.~\ref{fig:linescan}(b) how individual defect classes contribute to the enhanced emission at the crease. Just as NMF decomposition of a spectrum image enabled the extraction of spatially correlated spectral components in Figs.~\ref{fig:nmf_pl} and ~\ref{fig:nmf_cl}, NMF decomposition of this linescan (illustrated in Figs. ~\ref{fig:linescan}(c) and ~\ref{fig:linescan}(e)) can be used to extract weak spatially correlated defect bands. CL component~2 in Fig. ~\ref{fig:linescan}(e) exhibits weak emission around 750-850~nm and correlated emission in the visible, and it is strongly enhanced on the crease as evident in Fig.~\ref{fig:linescan}(c). CL components~1 and 3 exhibit a moderate change in intensity near the crease, but they are not strongly positively correlated with the crease position and the SE signal like component 2.

The number of components in each NMF reconstruction was chosen by monitoring the reconstruction error as a function of component number.  The PL and CL NMF decompositions require a different number of components because of the different relaxation pathways available under electron and laser excitation. Similarly, a reduced number of components were required for the linescan because the reduced scan area introduced fewer spatially correlated components.

Critically, the NMF decompositions in Figs.~2-4 illustrate selective enhancement of certain spectral components rather than uniformly enhanced luminescence at the creases.  Thus, the enhancement is not simply a result of increased interaction volume at the crease (which would uniformly enhance all unsaturated defect luminescence). Instead, it is logical to consider the effect of strain at the crease on defect luminescence. Following He-ion bombardment, large populations of both \VBn{} and \VN{} (including various charge states) are expected to be present in the material with roughly comparable densities\cite{Gu2021}.  According to previous density functional theory (DFT) calculations, the creased regions attract both \VBn{} and \VN{} defect species, as the formation energies of these defects are lowered with increasing curvature of hBN\cite{Zobelli2006}.

Here, we performed a series of DFT calculations of strained \VB{} defects to better understand the influence of strain on the \VB{} emission at the creases.  The calculations were performed with the Quantum Espresso package\cite{QE} with the PBE\cite{PBE1,PBE2} functional and ultrasoft GBRV\cite{GBRV} pseudopotentials.  Total energies were well converged with a planewave energy cutoff of 120 Ry.  The \VB{} structure was represented in 239 atom periodic supercells with a 2x2x2 k-point grid.  Relaxation of the local atomic structure was performed with the BFGS algorithm, with convergence tolerances of 1 meV for the total energy and 0.02 eV/\AA{} for forces. All calculations were spin polarized, consistent with the triplet ground state of \VB{}. 


Though both species are attracted to creases, a strong asymmetry exists between the mobility of \VBn{} and \VN{}.  The migration barrier of \VN{} is very large, effectively inhibiting its motion, while the barrier for \VBn{} migration is substantially lower\cite{Zobelli2007}, enabling much greater mobility.  The mobile \VBn{} and \VB{} species are therefore predicted to preferentially collect at creases in hBN, resulting in enhanced PL and CL in the 750-850 nm range at the creases driven by enhanced local populations of \VB{} defects in those regions.  An increased density of \VB{} defects at creases is also consistent with the increased slope seen in the power dependent PL plotted in Fig.\ref{fig:summary_overview}(e). 

The local strain environment at the creases may also contribute to changes in \VB{} emission relative to the pristine and locally flat regions.  Computational studies of hBN nanotubes have shown that the tubes constrict along the axial direction relative to pure hBN\cite{Blase1994,Jhi2005,Chen2010}, indicating a roughly uniaxial compressive strain at the creases.  DFT calculations performed in this work for \VB{} defects under uniaxial compressive strain show a modest redshift in the primary emission line of \VB{} as the strain is increased. Based on these calculations, we find that uniaxial strain of approximately 2.5\% is sufficient to explain the observed 30~nm redshift in the PL spectra at the creases relative to the unstrained regions. The assumption of uniaxial compressive strain is likely an over-simplification; because this is a relatively thick flake, it's likely that the strain environment varies from the top to the bottom of the flake, and we were not able to experimentally quantify the depth profile of the \VB{} defects in the flake.  


Notably, the NMF algorithm used here is an example of ``blind'' spectral unmixing.  In other words, no assumptions were made about the structure of each spectral component. The presence of at least two electronic transitions in a given NMF component does not necessarily suggest that a single defect is responsible for multiple transitions, but it does suggest that all electronic transitions in each component are spatially correlated with one another. The bands centered near 800 nm in PL component 6 and CL component 5 in Figs.~\ref{fig:nmf_pl} and ~\ref{fig:nmf_cl} respectively are consistent with \VB{} luminescence, and the ODMR spectrum shown in Fig.~\ref{fig:summary_overview}(c) corroborates this assignment.  However, most models of \VB{} luminescence suggest that it has only a single optically active transition centered near 800 nm\cite{ivady2020ab}.  Because helium irradiation should create comparable densities of \VBn{} and \VB{} and because these defects should exhibit comparable mobility, it seems reasonable to conclude that the transition at 570 nm in PL component 6, and the transitions at 550 nm and 400 nm in CL component 5 are a result of \VBn{} luminescence, consistent with calculations predicting \VBn{} luminescence at 2.3 eV and 2.8 eV\cite{huang2012defect} under the electron-hole recombination conditions that are possible in CL. Because PL component 2 exhibits a large shoulder at 800 nm and exhibits substantial strain enhancement at each of the creases, it similarly seems reasonable to assign PL component 2 to a higher energy \VBn{} transition with spatially correlated \VB{} luminescence.

The luminescence in PL components 4 and 5 and CL component 4 in Figs.~\ref{fig:nmf_pl} and ~\ref{fig:nmf_cl} respectively are not strongly correlated with the \VB{} band near 800 nm, and thus it seems unlikely that they are \VB{} luminescence bands.  Based on the understanding that helium irradiation should generate a moderate density of \VN{} defects with reduced mobility compared with \VB{} and \VBn{}, PL component 5 and CL component 4, which exhibit roughly uniform luminescence across the flake, can be assigned to \VN{} luminescence. CL component 4 also exhibits strong emission near 460 nm, which corresponds closely to the onset of photo-current observed in prior optical absorption measurements of hBN\cite{Remes2005}.  It is therefore possible to interpret this peak, which has broad spatial coverage over the sample, as being associated with the photo-ionization of \VN{} leading to the production of free carriers.

Despite the strong correlations between the PL and CL NMF decompositions, there are some visible distinctions between the spectral components. There are a few reasons for this: (1) most critically, PL microscopy relies on a below-bandgap excitation, but the electron beam excites electrons to the conduction band, resulting in the emergence of additional decay pathways that are not present in PL (and the PL map includes Raman modes that are not present in CL). (2) Many defect classes in hBN are sensitive to electron beam-excitation. While the electron beam doses that we used here did not visibly damage the hBN flake, it is possible that beam-induced damage introduced additional defects or quenched existing defects during measurement. (3) Cathodoluminescence microscopies of NV centers in diamond have shown that the electron beam converts the negatively charged NV center to the neutral NV center\cite{sola2019electron,feldman2018colossal}.  This beam-induced charge-state conversion substantially changes the defect photophysics.  Similar charge-state conversion likely occurs for the hBN defects probed here. 

Cathodoluminescence microscopies have not seen widespread use in the characterization of thin 2D materials because small interaction cross-sections limit the measurement efficiency. The spectral decomposition of CL spectrum images used here provides a more complete picture of the effect of flake morphology on correlated spectral components with spatial resolution of order 5 nm. A clear understanding of the nanophotonic properties of \VB{} defects in hBN is essential to the development of 2D quantum sensors. Atomically thin quantum sensors could offer improved sensitivity compared with bulk diamond-based sensors, and 2D materials are easier to integrate into devices.  However, \VB{} luminescence remains very weak compared with NV center luminescence.  The strain-induced funneling of \VB{} defects into intrinsic creases in hBN flakes that we have described here is one critical path toward addressing this obstacle. 

\section{Acknowledgments}
The authors would like to acknowledge Vladimir Shalaev for ODMR spectroscopy support. This research was sponsored by the U. S. Department of Energy, Office of Science, Basic Energy Sciences, Materials Sciences and Engineering Division. The PL and CL microscopy were performed at the Center for Nanophase Materials Sciences, which is a U.S. Department of Energy Office of Science User Facility. Sample preparation was supported by the DARPA QUEST Program and by the U.S. Department of Energy, Office of Science, National Quantum Information Science Research Centers, Quantum Science Center. Student support was provided by the U.S. Department of Energy, Office of Science, Office of Workforce Development for Teachers and Scientists, Office of Science Graduate Student Research (SCGSR) program and by NSF award DMR-1747426. The SCGSR program is administered by the Oak Ridge Institute for Science and Education for the DOE under contract number DE-SC0014664. This research used resources of the Compute and Data Environment for Science (CADES) at the Oak Ridge National Laboratory, which is supported by the Office of Science of the U.S. Department of Energy under Contract No. DE-AC05-00OR22725.

\providecommand{\latin}[1]{#1}
\makeatletter
\providecommand{\doi}
  {\begingroup\let\do\@makeother\dospecials
  \catcode`\{=1 \catcode`\}=2 \doi@aux}
\providecommand{\doi@aux}[1]{\endgroup\texttt{#1}}
\makeatother
\providecommand*\mcitethebibliography{\thebibliography}
\csname @ifundefined\endcsname{endmcitethebibliography}
  {\let\endmcitethebibliography\endthebibliography}{}

\end{document}